# Water-induced high-performance quantum-dot light-emitting diodes

**Authors:**

Wangxiao Jin[1]†, Siyu He[1]†, Xiuyuan Lu[1]†, Xitong Zhu[1], Dijiong Liu[1], Guolong Sun[1], Yanlei Hao[1], Xiaolin Yan[2]*, Yiran Yan[2], Longjia Wu[2,3], Xiongfeng Lin[3], Wenjun Hou[3], Weiran Cao[3], Chuan Liu[4], Xiaoci Liang[4], Yuan Gao[5], Yunzhou Deng[6], Feng Gao[7], & Yizheng Jin[1]*

**Affiliations:**

[1]State Key Laboratory of Silicon and Advanced Semiconductor Materials, Zhejiang Key Laboratory of Excited-State Energy Conversion and Energy Storage, Zhejiang University, Hangzhou 310058, China.

[2]TCL Corporate Research, 1001 Zhongshan Park Road, Nanshan District, Shenzhen 518067, China.

[3]Guangdong Juhua Research Institute of Advanced Display, Block A 7th Floor, No.11 Guangpu Middle Road, Huangpu District, Guangzhou 510663, China.

[4]State Key Laboratory of Optoelectronic Materials and Technologies, Guangdong Province Key Laboratory of Display Material and Technology, School of Electronics and Information Technology, Sun Yat-sen University, Guangzhou 510275, China.

[5]Najing Technology Corporation Ltd., Hangzhou, China.

[6]Cavendish Laboratory, University of Cambridge, Cambridge, UK.

[7]Department of Physics, Chemistry and Biology (IFM), Linköping University, Linköping, Sweden.

†These authors contribute equally to this work.

*Corresponding author: Prof. Yizheng Jin (yizhengjin@zju.edu.cn) and Dr. Xiaolin Yan (yan-xl@tcl.com).

**Solution-processed light-emitting diodes (LEDs) are appealing for their potential in the low-cost fabrication of large-area devices[1-3]. However, the limited performance of solution-processed blue LEDs, particularly their short operation lifetime[4-8], is hindering their practical use in display technologies. Here, we demonstrate that trace water in device — previously considered detrimental to most solution-processed LEDs—dramatically enhances the performance of quantum-dot LEDs (QLEDs). This breakthrough stems from our comprehensive mechanism investigations into the positive ageing phenomenon, a long-standing puzzle in the QLED field[9-11]. Our findings reveal that water passivation on the surface of electron-transport layers, which are composed of zinc-oxide-based nanoparticles, improves charge transport and enhances exciton radiative recombination during device operation. Combined with the advanced top-emitting architecture, our blue QLEDs achieve a high current efficiency of 35.5 cd $A^{-1}$, a blue index (colour coordinate corrected current efficiency) of over 470 cd $A^{-1}$ $CIE_y^{-1}$, and unprecedented stability, with an extrapolated $T_{95}$ lifetime (at an initial brightness of 1,000 cd $m^{-2}$) of 287 hours. Our work may inspire further exploration into surface passivation of nanocrystalline functional layers, critical for the advancement of emerging solution-processed optoelectronic and electronic devices.**

Electroluminescence (EL) of quantum dots (QDs) provides an attractive approach to exploit their efficient, stable and high-colour-purity luminescence properties[2,12-18]. Over the past decade, QLEDs have achieved significant advancements in both efficiency and operational lifetime[5,6,8,19-32]. Remarkably, most state-of-the-art QLEDs benefit from the so-called positive ageing effects, where devices show higher efficiencies and larger current densities after a period of shelf storage[5,6,23,27,28,30,32]. Previous research has traced the origin of positive ageing to the acidic resins used in device encapsulation[9,10,33]. It is hypothesized that the volatile acids released from the resins may passivate the surfaces of electron-transporting layers (ETLs) composed of ZnO-based nanoparticles, thereby contributing to performance enhancements. However, there is no direct evidence of acids penetrating the devices. This mechanism is plausible because the vigorous reaction between ZnO-based nanoparticles with acids may extend beyond mere surface passivation to etch the oxide nanocrystals or even convert them entirely into non-conductive zinc carboxylates[10]. Moreover, the progression of the acid-induced in-situ reactions results in unpredictable device performance improvements during initial shelf storage and subsequent deterioration after extended periods, both of which are untenable for commercial applications[9,10,33,34].

Here we reassess the molecular mechanism behind positive ageing using in-situ Fourier-transform infrared spectroscopy (FTIR, refer to Fig. 1a and Methods for details). We prepared a sample consisting of $CaF_2$/QDs/$Zn_{0.85}Mg_{0.15}O$ nanocrystals/patterned Au, encapsulated with an acrylic resin, which was identified as containing an additive of acrylic acid. A control sample was similarly prepared but encapsulated with an acid-free epoxy resin. Reflection-mode FTIR analyses of the Au-patterned area in both samples reveal the absence of C=C double bond stretching vibrations from acrylic acid (at 1,646 cm$^{-1}$, see also Extended data Fig. 1) and negligible changes in the intensities of the carboxylate absorption bands (asymmetric stretch and symmetric stretch at 1,595 and 1,430 cm$^{-1}$, respectively) (Fig. 1b). The most distinguishable changes observed between the two samples are the enhancements in the broad band at 3,000-3,600 cm$^{-1}$. These results suggest that the volatile acids do not diffuse into the electrode-patterned area of the oxide-nanocrystal layer. Instead, water molecules, in-situ generated by the reaction between acrylic acid and ZnO-based nanocrystals in the exposed area (without the top metal layer), readily infiltrate between the oxide nanocrystals in the electrode-patterned area (Fig. 1c). This hypothesis was confirmed by our in-situ FTIR analyses on the samples exposed in a controlled water (or deuterium water) atmosphere (Fig. 1b).

Our findings suggest that water, rather than acid, is primarily responsible for the positive ageing observed in QLEDs. Based on this insight, we developed a water treatment process (see Methods and Extended Data Fig. 2 for details). Briefly, we fabricated green QLEDs with a structure of ITO/poly(3,4-ethylenedioxythiophene):poly(styrenesulfonate) (PEDOT:PSS)/poly(*N*,*N'*-bis(4-butylphenyl-*N*,*N'*-bis(phenyl)-benzidine) (poly-TPD)/poly((9,9-dioctylfluorenyl-2,7-diyl)-*alt*-(9-(2-ethylhexyl)-carbazole-3,6-diyl)) (PF8Cz)/green CdSe/CdZnSe/ZnS (core/shell/shell) QDs (Extended Data Fig. 3)/$Zn_{0.85}Mg_{0.15}O$ nanocrystals/Ag[32]. The devices were then exposed to a controlled water atmosphere. An elevated temperature of 85 °C was applied to promote rapid diffusion of water molecules into the devices. Additionally, we employed a mixture of water and ionic liquid to control the partial pressure of water, preventing the condensation of liquid water which could damage the QLEDs. After the water treatment, the devices were encapsulated with acid-free epoxy resins.

Figure 2a-c shows that our water treatment process substantially improves the performance of green QLEDs. The device conductivity increases, with the current density at 4 V raising from 35.1 mA cm$^{-2}$ to 100.0 mA cm$^{-2}$ (Fig. 2a). The peak external quantum efficiency (EQE) of this device improves from 10.5% to 21.6% (Fig. 2b). Measurements across 27 devices yield an average peak EQE of 20.3% with a standard deviation of 0.8%, demonstrating excellent reproducibility (Fig. 2f, top). Remarkably, this treatment also boosts operational stability (Fig. 2c), extending the $T_{95}$ operation lifetime from 8 h at an initial brightness of 9,827 cd m$^{-2}$ to 100 h at an initial brightness of 14,653 cd m$^{-2}$. The $T_{95}$ at 1,000 cd m$^{-2}$ for the water-treated devices is estimated to be 16,900 h (with an acceleration factor of ~1.92), representing one of the most stable green QLEDs reported to date[6,30-32]. Given that D$_2$O molecules, similar to water, can diffuse into devices (Fig. 1b), we extended the water treatment process using D$_2$O. The results reveal comparable improvements in QLED performance, including higher EL efficiency, larger current density, and longer operational lifetime (Extended Data Fig. 4).

We highlight that our controlled water treatment process effectively prevents the progression of in-situ reactions caused by the acidic encapsulants, which continuously generate excess water by-products. Consequently, the water-treated QLEDs achieve long-term shelf stability, which is essential for commercial applications. As illustrated in Fig. 2d-e, all characteristics of a water-treated QLED, including current density-EQE (Fig. 2d) and voltage-current density-luminance (Fig. 2e), remain almost identical after 60 days of storage. Post-storage measurements on the 27 devices show an average peak EQE of 20.2% with a standard deviation of 0.7%, consistent with the initial values (Fig. 2f, bottom).

To further understand the microscopic correlation between the water treatment process and the enhanced device performance, we applied this treatment to a set of single-carrier devices with functional layers sandwiched between electrodes. Electrical measurements indicate that the electron-only device comprising Zn$_{0.85}$Mg$_{0.15}$O-nanocrystal films exhibits significantly higher current densities after the treatment (Fig. 3a), whereas no changes were observed in the hole-only devices comprising other functional layers (Extended Data Fig. 5). These results suggest that the water treatment predominantly affects the ETL of ZnO-based nanocrystals.

We carried out optical measurements on the water-treated Zn$_{0.85}$Mg$_{0.15}$O films. High-sensitivity photo-thermal deflection spectroscopy (PDS) reveals a reduction in intragap states after the water treatment[35,36] (Fig. 3b). Additionally, ultraviolet-visible spectroscopy (UV-Vis) analyses show a redshift in the absorption edges (Fig. 3b, inset), indicative of sintering or necking of the oxide nanocrystals. Together, the reduction of intragap states and the necking/sintering of the oxide nanocrystals shall enhance charge transport in the ETL films, thereby improving their conductance. Detailed analyses of the current density-voltage curves from the electron-only device corroborated this improvement (Extended Data Fig. 6). In the linear (Ohmic) regime with J = $A$V, the pre-factor $A$ for the water-treated device is found to be 33 times greater than that of the pristine device. The significantly lower resistance indicates a higher concentration of free carriers and increased carrier mobility in the water-treated oxide films. In the trap-filled limited space-charge limit current regime with J = $A'$V$^\alpha$, the pristine device exhibits a power factor $\alpha$ of voltage as 3.5, while the water-treatment device shows a power factor $\alpha$ of 2.7. The reduced power factor in the water-treated device suggests a decreased density of localized states, which typically act as defects below the conduction band edge[37]. Collectively, these results in both Ohmic and TFL-SCLC regimes support that water-treatment oxide ETLs are substantially more conductive and less defective, thus improving the conductance of the QLEDs.

We highlight that the improved ETLs, being less defective, enhance exciton radiative recombination during device operation. In our green QLEDs, the use of PF8Cz hole-transporting layers effectively mitigates electron leakage[6,32] (Extended Data Fig. 7), addressing one major efficiency-loss channel. Thus, we conducted in-situ EL-photoluminescence (PL) analyses to monitor the changes in exciton recombination efficiency (Fig. 3c). The results show that prior to the device turn-on, the PL efficiency of QDs in the water-treated devices is lower than that in the pristine devices. Remarkably, once the device is turned on, the PL efficiency of QDs in the water-treatment devices surpasses that of the pristine devices. These results collectively suggest the observed efficiency enhancement in the water-treated QLEDs can be primarily attributed to the enhanced exciton radiative recombination.

Finally, we extended the water treatment to top-emitting blue QLEDs. In general, top-emitting devices may offer enhanced current efficiency, high colour purity, and compatibility with panel-manufacturing processes, along with high aperture ratios, all of which benefit the overall design of emissive displays[38]. We selected blue CdZnSe/ZnSe/ZnS (core/shell/shell) QDs, featuring a PL peak at 473 nm with a full width at half-maximum of 15 nm (Extended Data Fig. 3). Due to the optimized microcavity structure of the top-emission devices (Fig. 4a), the resulting blue QLEDs demonstrate an EL peak at 474 nm with a narrower full width at half-maximum of 11 nm, corresponding to Commission Internationale de l'Eclairage (CIE) colour coordinates of (0.118, 0.075) (Fig. 4b). The maximum current efficiency of our blue QLED reaches 35.5 cd A$^{-1}$ (Fig. 4c). Consequently, the blue index—i.e., colour coordinate corrected current efficiency, a commonly used figure of merit for evaluating blue LEDs in the display industry[39-42]—of our blue QLED surpasses 470 cd A$^{-1}$ CIE$_y^{-1}$, outperforming all other solution-processed blue LEDs. Employing the relation $L_0^n T_{95}$ = constant and fitting an acceleration factor of $n$ = 1.69, the $T_{95}$ lifetime at an initial luminance of 1,000 cd m$^{-2}$ is predicted to be ~ 287 h (Fig. 4e). The $T_{50}$ lifetime at an initial luminance of 100 cd m$^{-2}$ is estimated to be 89,500 h (Extended Data Fig. 8). According to a comparison of blue index and operation lifetime (Extended Data Fig. 9 and Extended Data Table 1), our top-emitting blue QLEDs represent the best-performing solution-processed blue LEDs to date[4-8,43-50], comparable to state-of-the-art vacuum-deposited organic LEDs[41,42,51-54].

In summary, we have identified that water is the true molecular origin behind the long-standing puzzle of positive ageing in QLEDs. It is indeed interesting to see that water, often considered detrimental to most solution-processed LEDs, could be used to produce less defective and high-quality ZnO-based ETLs. Our straightforward water treatment method is effective in enhancing the efficiency, conductance, operational stability, and shelf stability of QLEDs. The unprecedented performance levels achieved by our blue QLEDs mark important progress towards large-area display applications.

# Figure Legends

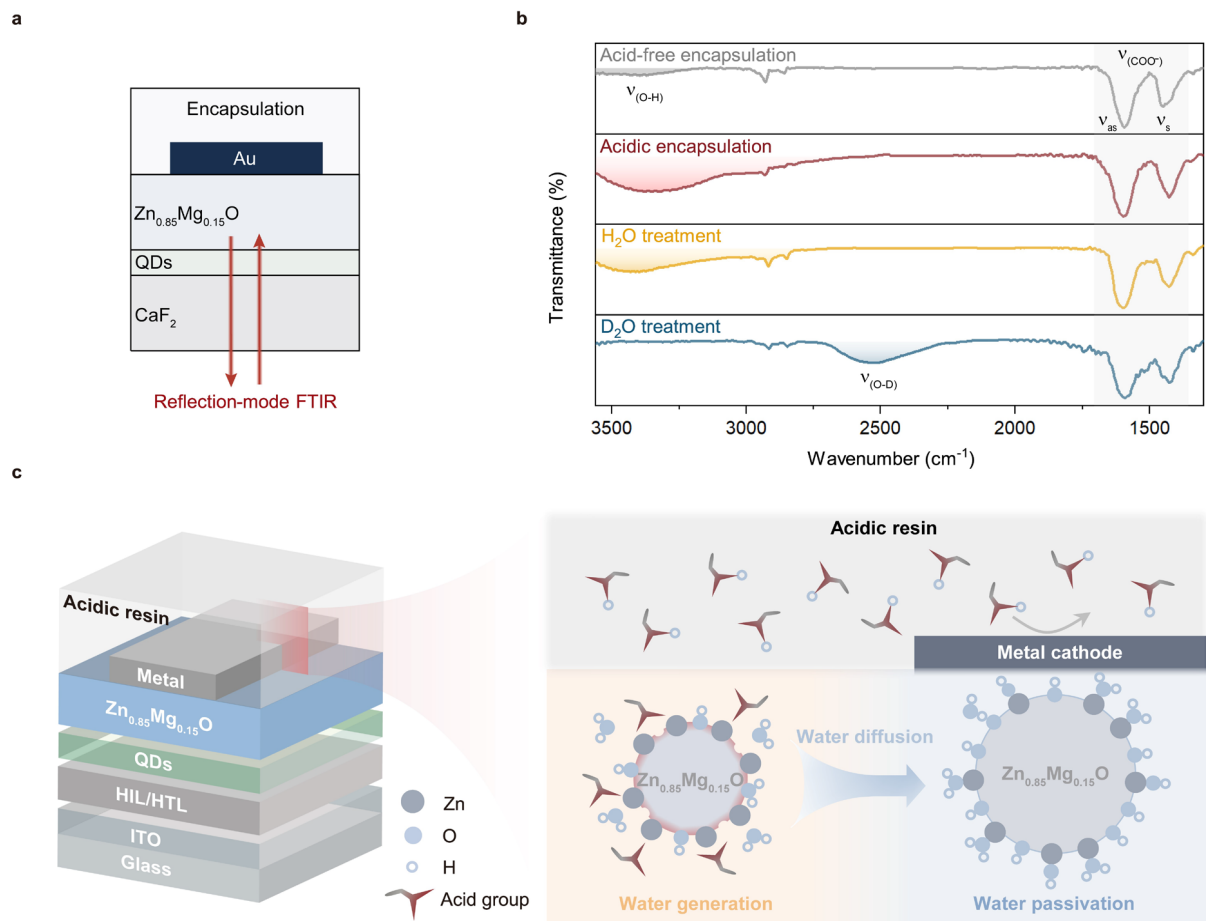

**Fig. 1 | Molecular mechanism responsible for positive ageing in QLEDs. a**, Schematic of the in-situ FTIR characterization. **b**, In-situ FTIR spectra of samples with different treatments. The $H_2O$- or $D_2O$-treated samples were encapsulated with acid-free epoxy resin. The vibration bands of 1,300–1,700 $cm^{-1}$, 2,300–2,700 $cm^{-1}$, and 3,000–3,600 $cm^{-1}$ correspond to the functional groups of -COO⁻, -OD, and -OH, respectively. **c**, Schematic illustration of the molecular mechanism behind positive ageing. The left side shows a QLED with acidic resin encapsulation. Exposed $Zn_{0.85}Mg_{0.15}O$ nanocrystals (without metal electrode coverage) are etched by volatile organic acids released from the resin, producing water. Subsequently, the in-situ generated water molecules, but not the volatile organic acids, diffuse into the device area and interact with the ZnO-based nanocrystals, inducing the positive ageing effects.

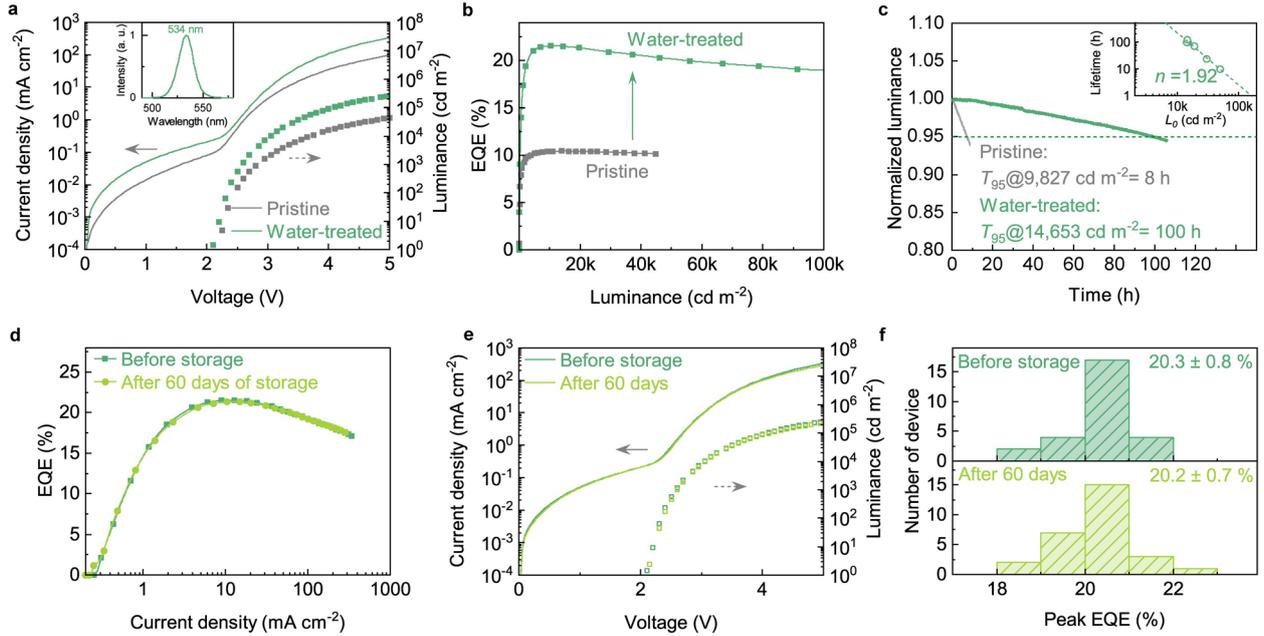

**Fig. 2 | Water treatment induced high-performance green QLEDs. a**, Current density-voltage-luminance characteristics of the pristine and the water-treated green QLEDs. The inset shows an EL spectrum at a current density of 100 mA m$^{-2}$. **b**, EQE-luminance curves of the pristine and the water-treated devices. **c**, Operational stability data of the pristine and the water-treated green QLEDs. The inset shows the $T_{95}$ lifetimes at different initial luminance ($L_0$) for the water-treated green QLEDs. The acceleration factor ($n$) is fitted to be ∼1.92 using the empirical relationship of $L_0^n T_{95}$ = constant. **d**, EQE-current density and **e**, current density-voltage-luminance characteristics of a water-treated device before and after 60 days of storage. **f**, Histograms of peak EQEs measured from 27 water-treated devices, before (top) and after (bottom) 60 days of storage.

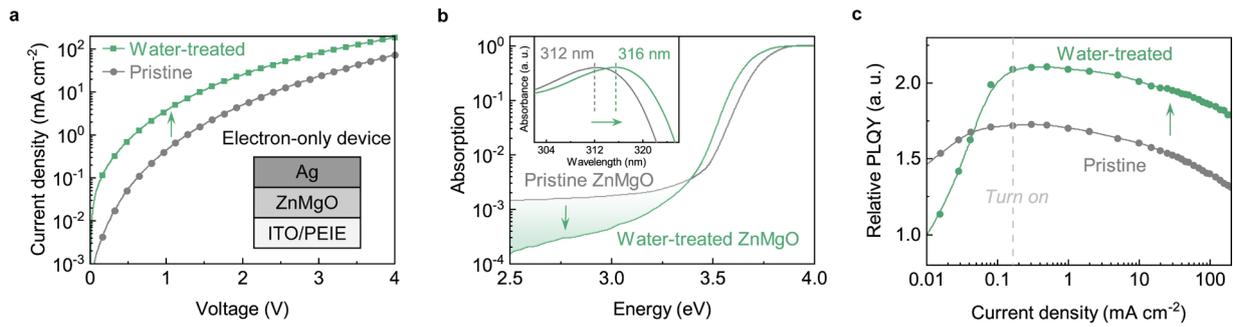

**Fig. 3 | Correlations between the water treatment and the improved device performance. a**, Current density-voltage characteristics of electron-only devices. The insert shows the device structure. The thickness of the $Zn_{0.85}Mg_{0.15}O$ layer is 50 nm, identical to that in QLEDs. **b**, PDS spectra of the $Zn_{0.85}Mg_{0.15}O$ films (deposited onto fused quartz substrates). The arrow indicates a decrease in parasitic absorptions of the water-treated $Zn_{0.85}Mg_{0.15}O$ films compared to those of the pristine films. The inset shows the UV-Vis absorption spectra of the $Zn_{0.85}Mg_{0.15}O$ films, with the arrow marking a redshift in the first exciton absorption peak after the water treatment. **c**, Relative PL efficiency of the QDs in the operational QLEDs.

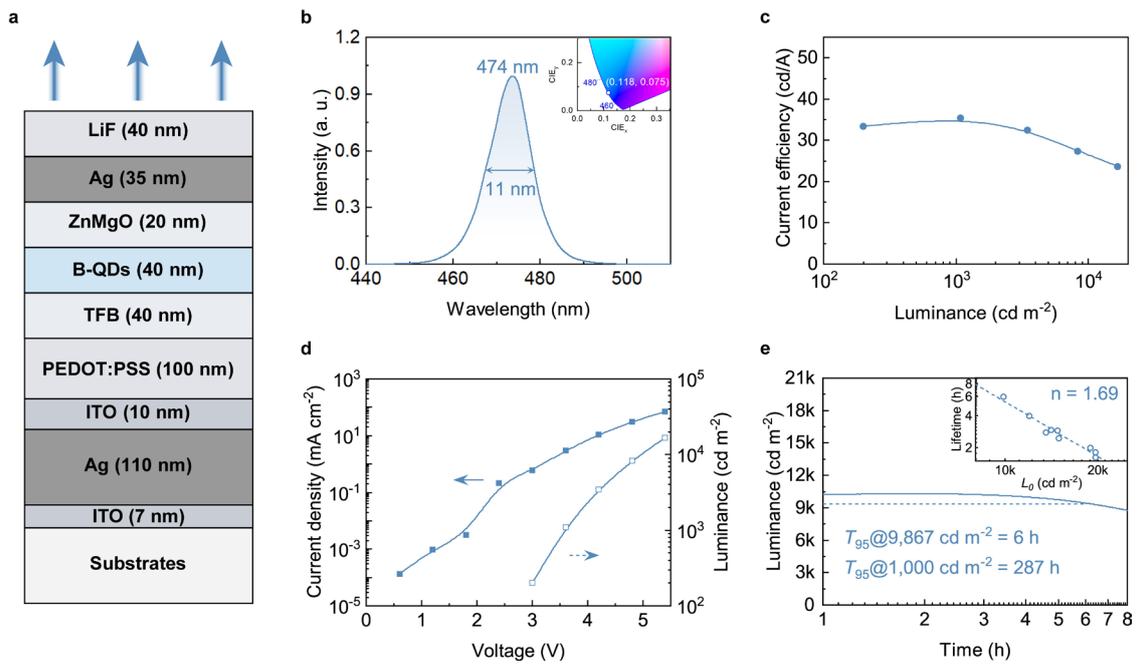

**Fig. 4 | Water treatment induced high-performance top-emitting blue QLEDs. a**, Device structure. **b**, EL spectrum and the corresponding CIE coordinates (inset). **c**, Current efficiency versus current density, and **d**, current density-voltage-luminance characteristics of a high-efficiency device. **e**, Stability data of the blue QLEDs. The $T_{95}$ lifetimes at various $L_0$ are shown in the inset. The acceleration factor ($n$) is fitted to be 1.69 by using $L_0^n T_{95}$ = constant.

**Extended data figure/table legends**

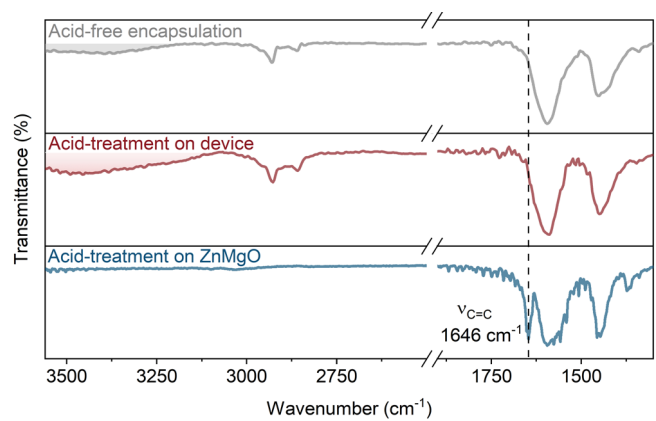

**Extended Data Fig. 1 | In-situ FTIR spectra of other control samples.** The sample treated with acrylic acid (red line) shows stronger -OH absorptions in the range of 3000–3600 cm$^{-1}$ and similar -COO$^-$ absorptions in the range of 1300–1700 cm$^{-1}$ compared to the acid-free encapsulated sample (grey line). The blue line represents the spectrum of a $Zn_{0.85}Mg_{0.15}O$ film pre-treated by acrylic acid vapour before the deposition of the Au electrode. The spectrum highlights a distinct absorption peak at 1646 cm$^{-1}$, which is attributed to the -C=C- stretching vibrations. The absence of the absorption peak at 1646 cm$^{-1}$ in the FTIR spectra of samples with acidic encapsulation (red line, Fig. 1b) or with the acrylic acid treatment (red line) indicates that the acid molecules cannot diffuse into the device area with the top patterned electrodes.

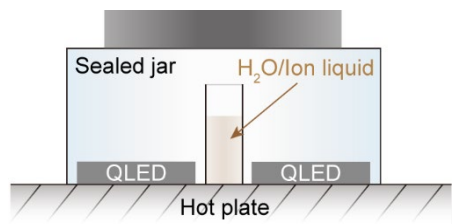

**Extended Data Fig. 2 | Schematic diagram of the water-treatment process.**

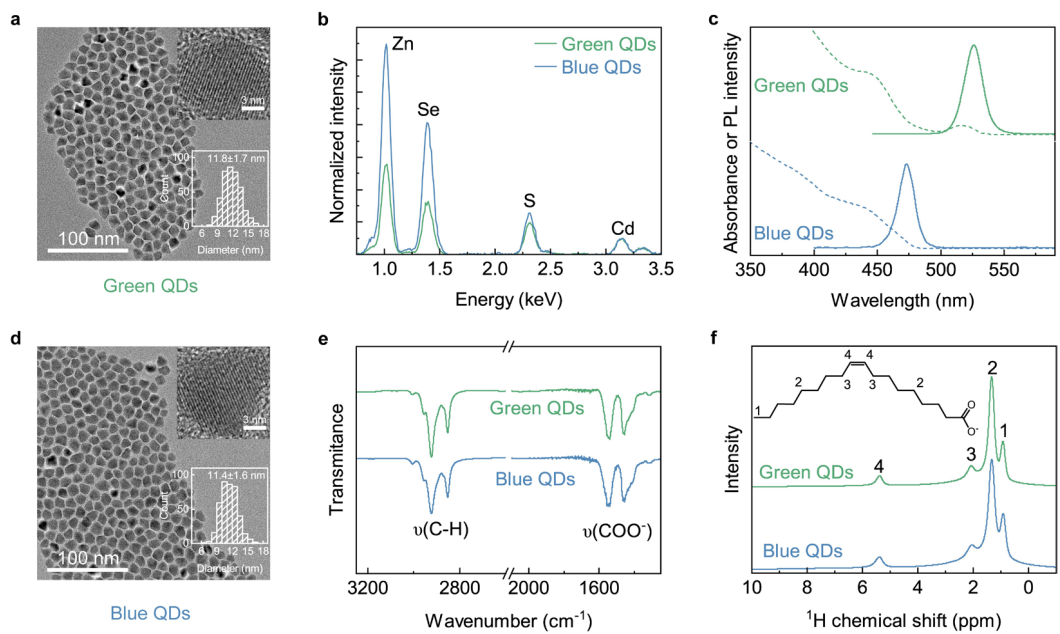

**Extended Data Fig. 3 | QD characterizations.** Transmission electron microscopy (TEM) images of (**a**) the green QDs and (**d**) the blue QDs. The insets are high-resolution TEM images of a green QD or a blue QD. The diameters of green QDs and blue QDs are measured to be 11.8 ± 1.7 nm and 11.4 ± 1.6 nm, respectively. **b**, Energy dispersive X-ray spectroscopy spectra extracted from elemental mapping results for Zn, Se, S, and Cd elements in the QDs. **c**, Absorption spectra of QD solutions (dashed lines) and PL spectra of QD films (solid lines). The peak wavelengths of green and blue QDs are 534 nm and 473 nm, respectively. **e**, FTIR spectra of the QD films showing characteristic absorption peaks of oleic carboxylate ligands. The absorption bands of 2,800–3,000 cm$^{-1}$ and 1,400–1,650 cm$^{-1}$ correspond to -C-H stretching vibrations and -COO$^-$ stretching vibrations, respectively. **f**, $^1$H solid-state nuclear magnetic resonance (NMR) spectra of the QDs showing characteristic chemical shift of oleic carboxylate ligands.

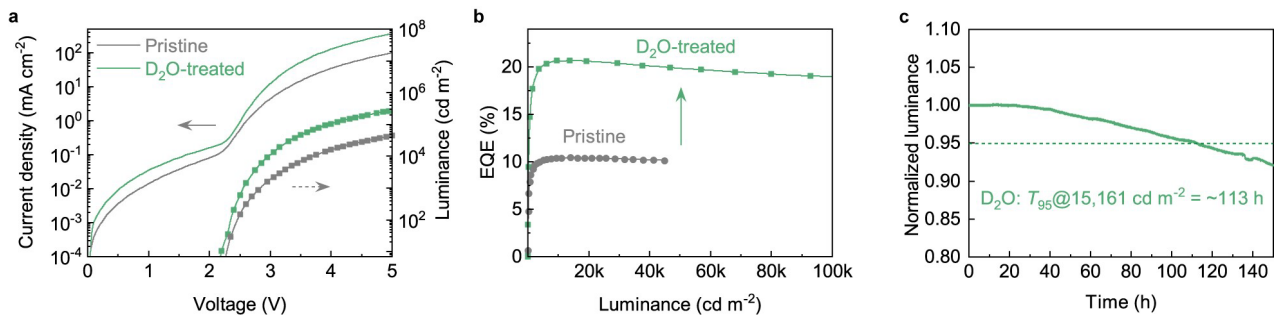

**Extended Data Fig. 4 | D₂O treatment induced high-performance green QLEDs.** (**a**) J-V-L characteristics, (**b**) EQE-luminance characteristics, and (**c**) operational stability of the D₂O-treated green QLEDs.

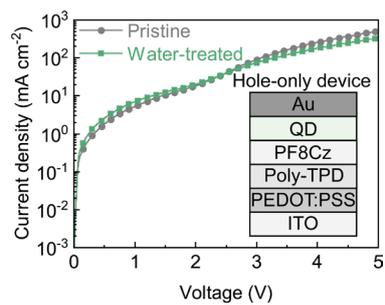

**Extended Data Fig. 5 | Current density-voltage characteristics of the hole-only devices.** The inset shows the device architecture, noting that the thicknesses of the functional layers are identical to those used in the QLEDs. The nearly identical J-V curves of the pristine and the water-treated devices suggest that water treatment has negligible effects on the transport properties of the QD layer, the hole-transport layer, and the hole-injection layer.

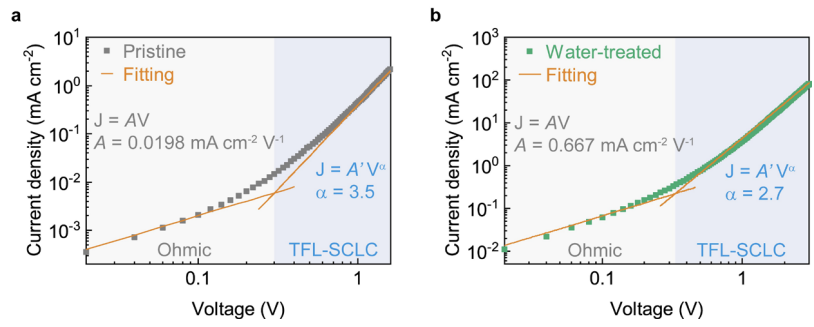

**Extended Data Fig. 6 | J-V characteristic curves and fittings for (a) the pristine and (b) the water-treated electron-only devices in a log-log scale.**

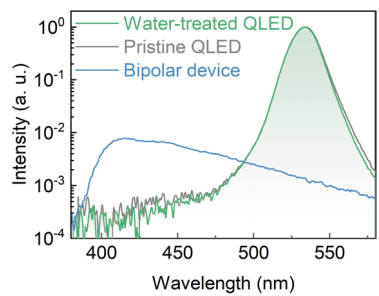

**Extended Data Fig. 7 | EL spectra of the green QLEDs and a bipolar device.** The structure of the bipolar device is ITO/PEDOT:PSS/PF8Cz/Zn$_{0.85}$Mg$_{0.15}$O/Ag. The EL emission at ~ 410 nm (blue line) is attributed to the electrical excitation of PF8Cz. The absence of PF8Cz emission in both water-treated and pristine QLEDs suggests effective suppression of electron leakage in these devices.

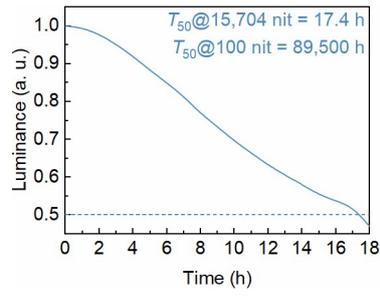

**Extended Data Fig. 8 | Measurements on $T_{50}$ lifetimes of a top-emission blue QLED.** The $T_{50}$ lifetime at the initial luminance of 100 cd m$^{-2}$ is estimated to be ~89,500 h by assuming an acceleration factor of ~1.69.

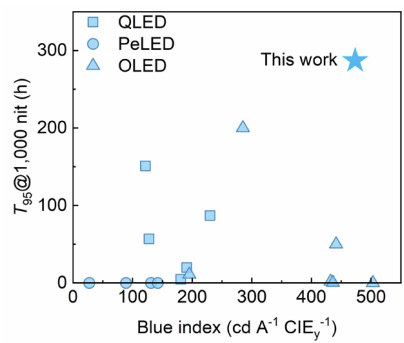

**Extended Data Fig. 9 | Performance Comparison of pure blue LEDs in the literature.** We note that devices with exceedingly short lifetimes or without available stability data are assigned a $T_{95}$ lifetime of 0 hours for clarity. Detailed data are provided in Extended Data Table 1.

**Extended Data Table 1 | Comparison of our devices with other solution-processed and organic blue LEDs**

| Device (Emissive materials) | Reference | $\lambda$ (nm) | CIE | Peak CE (cd A$^{-1}$) | Blue index (cd A$^{-1}$ CIE$_y^{-1}$) | $T_{95}$ (h) @1,000 nits | $T_{50}$ (h) @100 nits |
|---|---|---|---|---|---|---|---|
| **Solution-processed blue LEDs** | | | | | | | |
| QLED (CdZnSe/ZnSe/ZnS) | **This work** | **474** | **(0.118, 0.075)** | **35.5** | **473.3** | **287** | **89,500** |
| QLED (CdSe/ZnS) | *ACS A. M. I.* 2017[43] | 468 | (0.136, 0.078) | 14.1 | 180.8 | – | 47 |
| QLED (ZnTeSe/ZnSe/ZnS) | *Nature* 2020[5] | 460 | ~(0.136, 0.086) | ~16.4 | ~190.7 | ~20 | 15,850 |
| QLED (CdZnSe/ZnS) | *Nat. Photon.* 2022[6] | 479 | ~(0.100, 0.157) | ~20.0 | ~127.4 | 57 | 24,000 |
| QLED (ZnCdSe/ZnCdSeS/ZnS) | *Nat. Commun.* 2023[8] | 478 | (0.112, 0.125) | ~15.2 | ~121.6 | 151 | 50,206 |
| QLED (CdZnS/ZnS) | *Nat. Commun.* 2024[50] | 458 | (0.146, 0.040) | ~9.2 | ~230.0 | ~87 | 41,022 |
| PeLED (CsPbBr$_3$) | *Nat. Nanotechnol.* 2020[46] | ~478 | ~(0.091, 0.144) | ~12.8 | ~88.9 | – | ~0.3 |
| PeLED (CsPbBr$_3$) | *Adv. Mater.* 2021[47] | 470 | (0.130, 0.110) | ~3.0 | ~27.3 | – | 12 |
| PeLED (CsPbBr3) | *Nature* 2022[7] | 480 | (0.110, 0.130) | ~17.0 | ~130.8 | – | 2 |
| PeLED (CsPb(Br$_x$/Cl$_{1-x}$)$_3$) | *Sci. Adv.* 2024[49] | 475 | (0.105, 0.128) | ~18.2 | ~142.2 | – | – |
| s-TADF-OLED (5CzCN) | *ACS A. M. I.* 2018[44] | ~479 | (0.170, 0.310) | 52.3 | 168.7 | – | – |
| s-PH-OLED (Firpic) | *Sci. Adv.* 2016[4] | ~474 | (0.146, 0.372) | 38.7 | 104.0 | – | – |
| s-PH-OLED (FIr6) | *Adv. Opti. Mater.* 2019[45] | ~460 | (0.150, 0.250) | 41.9 | 167.6 | – | – |
| s-PH-OLED (mer-Ir(ptbbp)$_3$) | *Adv. Opti. Mater.* 2021[48] | 443 | (0.156, 0.102) | 22.4 | 219.6 | – | – |
| **Vacuum-deposited blue OLEDs** | | | | | | | |
| HF-OLED ($v$-DABNA) | *Nat. Photon.* 2021[51] | 470 | (0.150, 0.200) | 39.0 | 195.0 | 11 | – |
| TADF-OLED ($v$-DABNA) | *Nat. Photon.* 2021[41] | 473 | (0.120, 0.090) | 38.9 | 432.2 | ~2 | 6,100 |
| TADF-OLED (TBE02) | *Sci. Adv.* 2022[42] | ~463 | (-, 0.058) | 25.6 | 441.4 | ~50 | – |
| TADF-OLED ($v$-DABNA) | *Nat. Commun.* 2023[52] | 465 | (0.130, 0.100) | 50.3 | 503.0 | – | – |
| TSF-OLED ($v$-DABNA) | *Nat. Photon.* 2024[53] | 468 | (0.150, 0.200) | ~57.0 | 285.0 | ~200 | – |
| TADF-OLED (DPA-B4) | *Nat. Photon.* 2024[54] | 458 | (0.141, 0.050) | 21.8 | 436.0 | ~0.3 | 418 |

This table does not include tandem LEDs.

*, approximated values extracted from the figures in the literature.

–, values not available.

TADF: thermally activated delayed fluorescence; PH: phosphorescent; HF: hyperfluorescence; TSF: TADF-sensitized fluorescence.